\documentclass[aps,prb,preprint,superscriptaddress,showpacs]{revtex4}
\usepackage{amsmath}
\usepackage{bm}
\usepackage{graphicx}

\begin{document}

\title{Persistent current magnification in a double quantum-ring system}
\author{P. A. Orellana}
\affiliation{Departments de F\'{\i}sica, Universidad Cat\'{o}lica
del Norte, Casilla 1280, Antofagasta, Chile}
\author{M. Pacheco}
\affiliation{Departments de F\'{\i}sica, Universidad T\'ecnica F.
Santa Mar\'{\i}a, Casilla 110-V, Valpara\'{\i}so, Chile}

\begin{abstract}
The electronic transport in a system of two quantum rings side-coupled to a quantum wire
is studied via a single-band tunneling tight-binding Hamiltonian. We derived analytical
expressions for the conductance, density of states and the persistent current when the
rings are threaded by magnetic fluxes. We  found a clear manifestation of the presence of
bound states in each one of those physical quantities when either the flux difference or
the sum of the fluxes are zero or integer multiples of a quantum of flux. These bound
states play an important role in the magnification of the persistent current in the
rings. We also found that the persistent current keeps a large amplitude even for strong
ring-wire coupling.
\end{abstract}

\maketitle
\section{Introduction}

Electronic transport through quantum rings structures has become
the subject of active research during the last years.  Interesting
quantum interference phenomena have been predicted and measured in
these mesoscopic systems in presence of a magnetic flux, such as
the Aharonov-Bohm oscillations in the conductance, persistent
currents\cite{Chandra,Mailly,Keyser} and Fano antiresonances
\cite{damato,pedro2}. Also, optical spectroscopy measurements have
allowed a determination of the energy spectra of closed
semiconducting rings\cite{Lorke}. Recently, F\"uhrer\cite{Fuhrer}
reported magnetotransport experiments on closed rings showing the
Aharonov-Bohm effect on the energy spectra.

Since the persistent currents in isolated mesoscopic rings were
predicted \cite{Buttiker1} and posteriorly verified experimentally
\cite{Levy}, there have been many studied on the single rings and
coupled-rings systems. The persistent currents also have been
reported in open systems. In these systems, the persistent current
magnification effect have been predicted for several
authors\cite{pareek,yi,bando}. For instance Pareek and Jayannavar
\cite{pareek}reported persistent current magnification in two
mesoscopic rings connected to an electron reservoir, Yi et
al.\cite{yi} find giant persistent currents in open Aharonov-Bohm
rings and Bandopadhyay et al. \cite{bando} obtain persistent
current magnification in multichannel mesoscopic ring.

A mesoscopic ring coupled to a reservoir was discussed
theoretically a long time ago by B\"uttiker\cite{Buttiker}, in
which the reservoir acts as a source of electrons and an inelastic
scatterer. Takai and Otha\cite{Takai} considered the case where
the magnetic flux and an electrostatic potential were applied
simultaneously. The occurrence of persistent currents and the
behavior of the electric conductance along a normal metal loop
connected to two electron reservoirs was also discussed
previously\cite{Jayannavar}. Recently Wunsch et al. \cite{wunsch}
find the development of long living states in a single ring
coupled to a reservoir and discuss their effect in the persistent
currents and the relation with the Dicke effect in
optics.\cite{dicke}

In a previous paper (Ref \onlinecite{pedro2}) we investigate the conductance and the
persistent current of a mesoscopic quantum ring attached to a perfect quantum wire in
presence of a magnetic field. We show that the system develops an oscillating band with
resonances (perfect transmission) and antiresonances (perfect reflection). In addition,
an odd-even parity of the number of sites of the ring was found. With the advances in the
fabrication techniques of nanostructures, it is natural to propose and study more complex
mesoscopic ring-like structures, as a double quantum ring, and to develop a theoretical
quantum mechanical description for understanding such these new family of systems. In the
present work we address to the study of the transport properties of a system composed by
two quantum rings side attached to a quantum wire. We found persistent current
magnification in the rings due to the formation of bound states in the rings when the
magnetic flux difference is an integer number of the quantum of flux.

\section{Model}

\begin{figure}[t!]
\centerline{\includegraphics[width=7cm,angle=0]{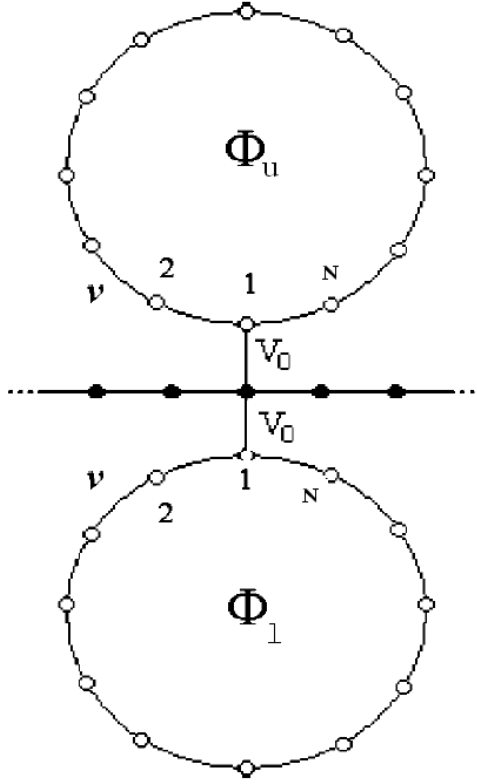}}
\caption{Schematic view of the two quantum ring attached to
quantum wire.} \label{fig1}
\end{figure}

The system under consideration is formed by two N-sites quantum
rings connected by tunnel coupling to a quantum wire waveguide of
infinity length, as shown schematically  in Fig.~\ref{fig1}. The
 rings are threaded by a magnetic field flux.
The full system is modeled by a single-band tight-binding
Hamiltonian within a noninteracting picture, that can be written
as
\begin{equation}
H=H_{W}+H_{R}+H_{WR},
\end{equation}
with
\begin{eqnarray}
H_{W}&=&-v\sum_{<i\neq j>}\,(c_{i}^{\dagger }c_{j}+c_{i}c_{j}^{\dagger }),
\nonumber \\
H_{R}&=&\sum_{n=1,\alpha=u,l}^{N}\varepsilon
_{n}^{\alpha}d_{n,\alpha}^{\dagger
}d_{n,\alpha} \nonumber \\
H_{WR}&=&-V_{0}\sum_{n=1,\alpha=u,l}^{N}(d_{n,\alpha}^{\dagger}c_{0}+c_{0}^{\dagger}d_{n,\alpha}),
\end{eqnarray}

\noindent where $c_{i}^{\dagger}$  is the creation operator of an electron at site $i$ of
the wire, and $d_{n,\alpha}^{\dagger}$ is the corresponding operator of an electron in
the state $n$ of the upper ($\alpha=u$) or lower ($\alpha=\ell$) ring. The wire
site-energy is assumed equal to zero and the hopping energies for wire and rings are
taken to be equal to $v$, whereas $V_{0}$ couples both systems. In the nearest neighbors
approximation the energy of the state $n$ of the ring $\alpha$ is
$\varepsilon_{n}^{\alpha}=2v cos(2\pi (n+\varphi_{\alpha})/N)$ where
$\varphi_{\alpha}=\Phi_{\alpha}/\Phi _{0}$, is the magnetic flux in unit of elemental
quantum flux ($\Phi _{0}=h/e$).

The linear conductance can be obtained from Landauer formula at
zero temperature:
\begin{equation}
\mathcal{G}=\frac{2e^2}{h}T(\omega=\varepsilon_F)
\end{equation}
\noindent where $T(\omega)$ is the transmission probability, given
by,

\begin{equation}
T(\omega)=\frac{2\Gamma_L(\omega)\Gamma_R(\omega)}{\Gamma_L(\omega)+\Gamma_R(\omega)}\Im
m[G^W_{0}],
\end{equation}
\noindent here $G^W_{0}$ is the Green's function of the site "0" of the wire, the
$\Gamma_{L(R)}$ is the coupling of the site "0" with the  left (right) side of the wire.

By using a Dyson equation $G=g+gH_{WR}G$ we calculate the Green's function of the site
$0$ of the quantum wire coupled to the rings, obtaining the following expression:
\begin{equation}
G_{0}^{W}=\frac{i}{2v\sin(k)}\,\,\frac{1}{1-i\gamma (Q_1+Q_2)},
\end{equation}
\noindent where $\gamma=\pi V_0^2/2v\sin(ka)$ ($ka=\arccos(-\omega/2v)$) and $Q_{\alpha}$
is given by the following equation,
\begin{equation}
Q_{\alpha}=\sum_{n=1}^{N} \frac{1}{\omega -
\epsilon_{n}^{\alpha}}=\frac{1}{2v}\frac{\sin(Nka)}{\sin(ka)}[\cos(2\pi
\varphi_{\alpha})-\cos(Nka)]^{-1}.
\end{equation}
\noindent By considering symmetric couplings
$\Gamma_L=\Gamma_R=\Gamma(\omega)=2v\sin(ka)$ and equations 3-5,
we evaluate the linear conductance,
\begin{equation}
\mathcal{G}=\frac{2e^2}{h}\Gamma(\omega)\Im m[G^W_{0}(\omega)]
\big |_{\omega=\varepsilon_F}=\frac{2e^2}{h} \frac{1}
{1+\gamma^2(Q_{1}+Q_{2})^2}\big |_{\omega=\varepsilon_F}.
\end{equation}

The density of states of the quantum rings can give us a better understanding of the
transport properties of the system. To obtain it we calculate the spectral function
$S_{n}^{\alpha}$ from the imaginary part of the diagonal elements of the Green's
functions of the rings, $G^{\alpha}$ ($\alpha=u,l$), this is:

\begin{equation}
G_{n}^{\alpha}=g_{n}^{\alpha}+\frac{i\gamma |g_{n}^{\alpha}|^2}{1-i\gamma (Q_u+Q_l)},
\end{equation}
then,

\begin{eqnarray}
S_{n}^{\alpha}&=&\frac{1}{\pi}\frac{\gamma}{(\omega-\varepsilon_{n}^{\alpha})^2}\,\,\frac{1}{1+\gamma^2
(Q_u+Q_l)^2}.
\end{eqnarray}

\noindent The density of states in each ring is  obtained summing over all the states of
the corresponding ring:

\begin{equation}
\rho_{\alpha}(\omega)=\sum_{n=1}^{N}S_{n}^{\alpha}(\omega)=\frac{\gamma}{\pi}\,\,
\frac{(-\partial Q_{\alpha}/\partial \omega)}{1+\gamma^2
(Q_u+Q_l)^2},
\end{equation}

In the presence of a magnetic flux threading the rings, a
persistent current is generated through the circular system. The
persistent current density $j_{\alpha}$ in the ring $\alpha$, in
the energy interval $d\omega $ around $\omega$, is given by

\begin{equation}
j_{\alpha}(\omega)=\frac{e}{h\pi}\sum_{n=1}^{N}\big (-\frac{\partial
\varepsilon_{n}^{\alpha}}{\partial\varphi_{\alpha}}\big)S_{n}^{\alpha}(\omega)
 =\frac{e\gamma}{h \pi}\,\,\frac{(-\partial Q_{\alpha}/\partial
\varphi_{\alpha})}{1+\gamma^2 (Q_u+Q_l)^2}. \label{pdc1}
\end{equation}

The total persistent current in each ring $\alpha$ is obtained integrating the persistent
current density over $\omega$,  we can write,

\begin{equation}
I_{\alpha}=\int j_{\alpha}(\omega)f(\omega)d\omega
\end{equation}

\noindent where $j_{\alpha}$ is given by Eq. \ref{pdc1} and $f(\omega)$ is the
Fermi-Dirac distribution. In order to study the behavior of the above physical quantities
as a function of the relevant flux parameters, we adopt new phases defined in terms of
the Aharonov-Bohm phases $\varphi_{\alpha}$. An added phase
$\tilde{\varphi}=(\varphi_{u}+\varphi_l)$, and a difference phase $\delta
\varphi=(\varphi_u-\varphi_{l})$. Introducing the following definitions,
\begin{eqnarray}
x&\equiv&\cos(\pi \tilde{\varphi})\cos(\pi
\delta \varphi)-\cos(Nka),\nonumber \\
y&\equiv&\sin(\pi \tilde{\varphi})\sin(\pi \delta \varphi),\nonumber \\
\beta&\equiv&\frac{1}{2v}\frac{\sin(Nka)}{\sin(ka)},\label{beta}
\end{eqnarray}
\noindent the conductance can be written as,
\begin{equation}
\mathcal{G}=\frac{2e^2}{h} \frac{[(x+y)(x-y)]^2} {[(x+y)(x-y)]^2+ (\gamma\beta)^2x^2}
\big|_{\omega=\varepsilon_F}. \label{con2}
\end{equation}
From these equations we note that the conductance vanishes each
time that $(x+y)=\cos(2\pi\varphi_l)-\cos(Nka)$ or
$(x-y)=\cos(2\pi\varphi_u)-\cos(Nka)$ are zero. Therefore the
conductance will present antiresonances for energies corresponding
exactly to the energy spectrum of the isolate rings. On the other
hand, the resonances in the conductance are obtained when $\beta$
or $x$ becomes null, except when $y$ is equal to zero which occurs
when $\tilde{\varphi}$ or $\delta \varphi$, are zero or integer
values.

Similarly the density of states (DOS) and persistent current
density (PCD) can be written in terms of flux parameters as,

\begin{eqnarray}
\rho_{\alpha}(\omega)&=& \frac{N\gamma (x\pm
y)^2}{\sin(ka)^2}\frac{[1-\cos(Nka)\cos(\pi(\tilde{\varphi}\pm\delta\varphi))-
(2v\beta/N)(x\mp y)]}{(x^2-y^2)^2+(2\gamma\beta)^2x^2} \label{dos2}
\end{eqnarray}

\begin{eqnarray}
j_{\alpha}(\omega)&=& -\frac{2e}{h}
\frac{v\sin(\pi(\tilde{\varphi}\pm\delta\varphi))(x\pm y)^2
 }{(x^2-y^2)^2+(2\gamma\beta)^2x^2} \gamma\beta. \label{PCD2}
\end{eqnarray}

\noindent In the above expressions the upper(lower) sign corresponds to $\alpha=u(l)$.
These analytical results permit us to analyze its behavior as a function of the different
relevant parameters of the system. We can see that both functions oscillate with the
energy and the magnetic flux parameters and that for each ring, DOS and PCD vanishes at
the eigenenergies corresponding to the other isolated ring, except when $y=0$. In
particular, it can be observed that the PCD in both rings vanishes for the zeroes of
$\beta$ (Ec. \ref{beta}), i.e. $\omega=-2v\cos(\pi j/N)$ ($j=1..N$). Furthermore, the
maxima amplitude of the DOS and PCD can be obtained from the minimum of the common
denominator in the equations \ref{dos2} and \ref{PCD2}, i.e.
$(x^2-y^2)^2+(2\gamma\beta)^2x^2$.

  In what follow  we present results for the conductance, density of
states, and  persistent current of a double ring system of $N=10$ sites, coupled through
a wire. Energies are measured in units of the hoping parameter $v$. Notice that all
studied physical quantities depend in a completely equivalent form of the flux parameters
$\tilde{\varphi}$ and $\delta \varphi$, therefore its can be analyzed as a function of
any of them and the wire-ring coupling energy $\gamma$. For given set of parameters the
energy spectrum consists of a superposition of quasi-bound states reminiscent of the
corresponding localized spectrum of the isolated rings.

\begin{figure}[b!]
\centerline{\includegraphics[angle=-90,scale=0.5]{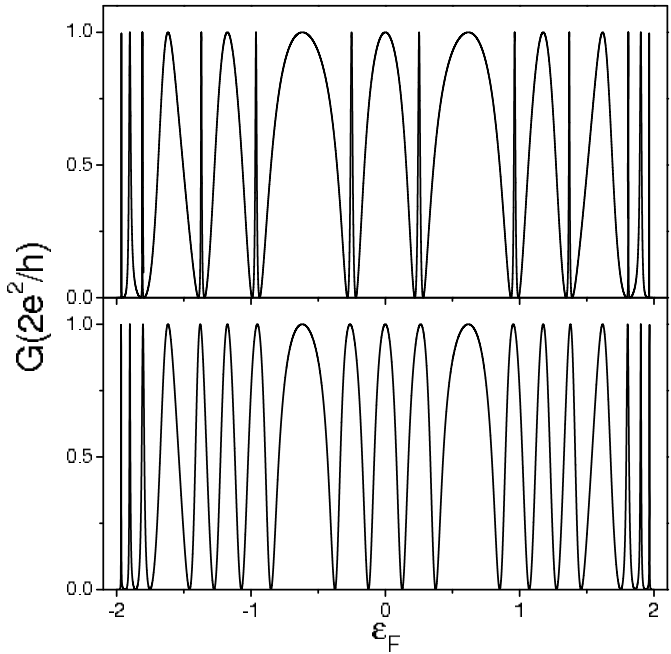}}
\caption{Linear conductance  as a function of Fermi energy
measured in units of the hoping parameter $v$, for $\gamma=0.5v$,
$\tilde{\varphi}=0.6$ and for a difference of fluxes between both
rings of $\delta \varphi=0.1$ (top) and $\delta \varphi=0.2$
(bottom).} \label{fig2}
\end{figure}

 Figure 2 displays the
linear conductance versus the Fermi energy for a coupling energy
$\gamma=0.5$ and parameters of magnetic flux given by
$\tilde{\varphi}=0.6$ and $\delta \varphi=0.1 $ (upper layer) and
$\delta \varphi=0.2$ (lower layer). As expected from the
analytical expression (Eq. \ref{beta}) the linear conductance
presents a series of resonances and antiresonances as a function
of the Fermi energy. The zeroes in the conductance represent
exactly the superposition of the spectrum of each isolated ring
 $\omega=-2v\cos(2\pi(n+\varphi_{u(\ell)})/N)$.
 Notice that when the difference between both fluxes decreases the
shape of the resonances changes drastically and the conductance becomes composed of broad
and narrow peaks. When the magnetic fluxes coincide the isolated-ring spectra are
degenerated and the narrow peaks are completely suppressed from the conductance. We will
analyze in detail this situation below.

\begin{figure}[]
\centerline{\includegraphics[angle=-90,scale=0.5]{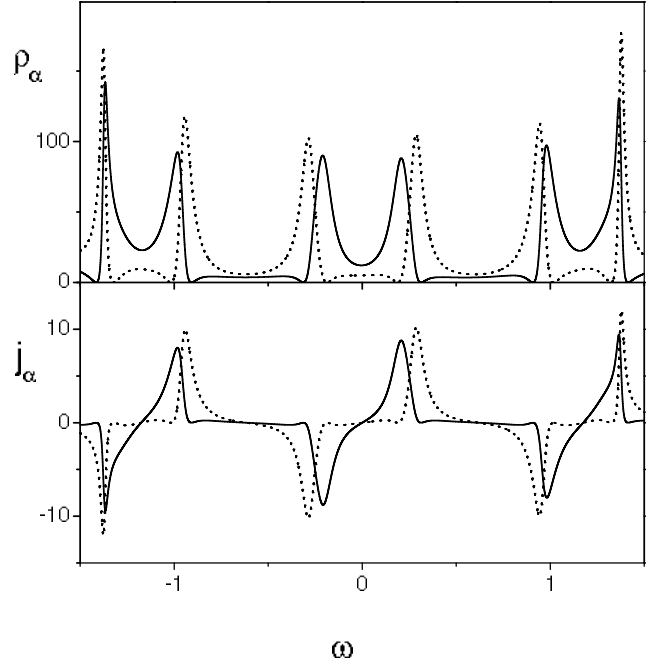}}
\caption{Density of states (top) and persistent density current(in
units of $v$), for $\tilde{\varphi}=0.6$, $\delta \varphi=0.1$ and
$\gamma=0.5$. The solid and dotted lines indicate upper ring  and
lower ring respectively.} \label{fig3}
\end{figure}

In figure 3 we have plotted the density of states (upper plot) and
the persistent current density (lower plot) as a function of
 energy $\omega$, for  $\gamma=0.5$ and flux
parameters given by $\tilde{\varphi}=0.6$ and $\delta \varphi=0.1$. The solid and dotted
lines indicate upper ring  and lower ring respectively. We can see the expected structure
of maxima and minima of DOS and PCD localized in the energy values defined by the
spectrum of each isolated ring. The wire-rings coupling energy determines the mixing
between the states of the two rings and in consequence the width of the peaks in the DOS
and PCD.

\begin{figure}[]
\vspace{2cm}
\centerline{\includegraphics[angle=-90,scale=0.5]{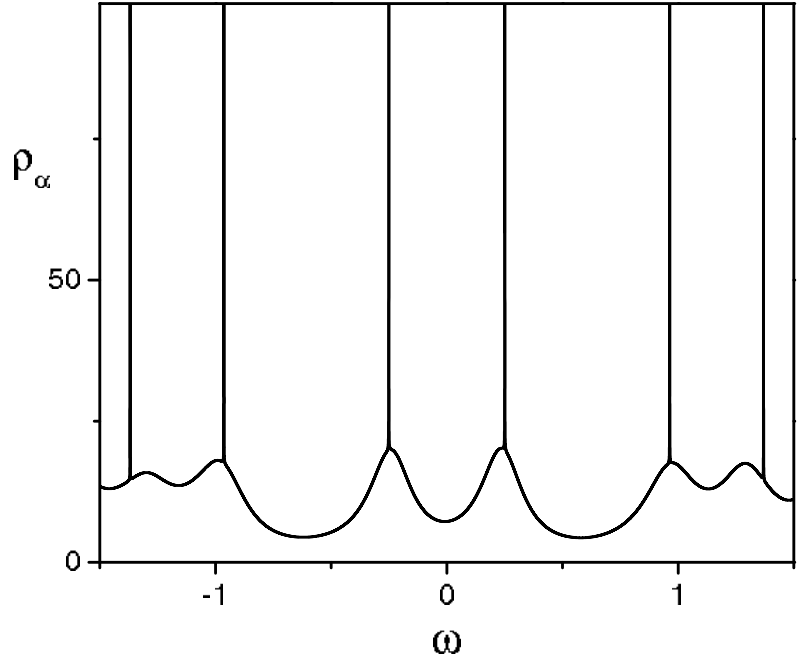}}
\caption{Density of states as a function of energy (in units of
$v$), for $\tilde{\varphi}=0.6$, $\delta \varphi=0.0$ and
$\gamma=0.5$. } \label{fig4}
\end{figure}

An interesting situation appears when the energy spectrum of both rings becomes
degenerated. This occurs when  the magnetic fluxes threading the rings are equals or they
are integer multiples one of each other. This is $\tilde{\varphi}=M$ or $\delta
\varphi=M$, with $M=0,\pm1,\pm2..$.  For this case $y\equiv\sin(\pi
\tilde{\varphi})\sin(\pi \delta \varphi)=0$, and we obtain,

\begin{eqnarray}
\mathcal{G}&=&\frac{2e^2}{h}\frac{x^2}{x^2+4\gamma^2},\nonumber\\
\rho_{\alpha}(\omega)&=&\rho_s(\omega)+\sum_{n=1}^{N}\delta(\omega-\varepsilon_n),\nonumber\\
j_{\alpha}(\omega)&=&j_s(\omega) +
\frac{e}{h}\sum_{n=1}^{N}\big(-\frac{\partial
\varepsilon_{n}}{\partial\tilde{\varphi}}\big)\delta(\omega-\varepsilon_n).
\end{eqnarray}
\noindent where $\rho_s$ and $j_s$ are given by

\begin{eqnarray}
\rho_{s}(\omega)&=&
\frac{N\gamma}{\sin(ka)^2}\frac{(1-\cos(Nka)\cos(\pi\tilde{\varphi})-(2v\beta/N)x
)}{x^2+4\gamma^2}\nonumber \\
j_{s}(\omega)&=& -\frac{e}{h}
\frac{2\gamma\sin(\pi\tilde{\varphi})}{x^2+4\gamma^2}.\label{pcs}
\end{eqnarray}

\noindent
 The analytical expression obtained for DOS clearly shows the presence of N
bound states  reminiscent of the spectrum of an isolated ring. It is clear that they do
not contribute to the conductance. Bound states appear superposed to the corresponding
spectrum of an effective single ring coupling to the quantum wire with coupling energy
equal to $2\gamma$. Figure 4 illustrates the behavior of the  DOS as a function of the
energy for $\delta \varphi=0.0$. Here $\gamma=0.5$ and $\tilde{\varphi}=0.6$. The DOS
shows the superposition of sharp and broad resonances centered at the same energies. It
is important to note that the above expression are valid only if $\gamma \ne 0$.

The presence of bound states induces divergences in the persistent
current density for energies corresponding to the eigenenergies of
the isolated rings. This affects the behavior of the total
persistent current inside the rings which is strongly enhanced for
certain values of the parameters as we will see below. The total
persistent current for the case of equals magnetic fluxes, ie.
$\delta \varphi=0.0$ can be written as:

\begin{eqnarray}
I_{\alpha}&=&\int j_{s}(\omega)f(\omega)d\omega +
\frac{e}{h}\sum_{n=1}^{N}\big(-\frac{\partial
\varepsilon_{n}}{\partial\tilde{\varphi}}\big)\Theta(\varepsilon_F-\varepsilon_n),
\nonumber \\
I_{\alpha}&=&I_s(\gamma,\tilde{\varphi})+I_b(\tilde{\varphi}).
\end{eqnarray}

\noindent where $j_{s}(\omega)$ is given by Eq. \ref{pcs}. Notice that the persistent
 current in each ring contains two contributions, one $I_s$ from a single ring with
 an effective coupling energy $2 \gamma$  and another one $I_b$,
 independent of the coupling parameter. In order to evaluate the total persistent current we put the Fermi energy
 at the center of the band i.e. $\varepsilon_F=0$ at zero temperature.
In figure 5 we illustrate the total persistent current in units of
the  persistent current of an isolated ring $I_0$, for different
values of $\gamma$. It can be observed that for small values of
$\gamma$ there is an notable enhancement of the total persistent
current in each ring. We observe that the amplitude of the
persistent current is nearly twice the corresponding amplitude of
the persistent current of an isolated ring $I_0$. This
magnification effect is associated with the presence of bound
states in the spectrum of the coupled-rings system.

Other interesting effect occurs for large values of $\gamma$. We observe that when
$\gamma$ increases, the amplitude of the persistent current tends to $I_0$, this is
contrary to the case of a single ring connected to reservoirs for which the amplitude of
the persistent current decays to zero for increasing $\gamma$. We show this effect in
figure 6 which displays the amplitude of the total persistent current $I_T$ and the
amplitude from the contributions $I_s$ and $I_b$ as a function of coupling parameter. We
see that for $\gamma$ sufficiently large the amplitude of $I_s$ tends to zero and the
persistent current tends to $I_0$.

\begin{figure}[]
\vspace{2cm} \centerline{\includegraphics[angle=-90,scale=0.5]{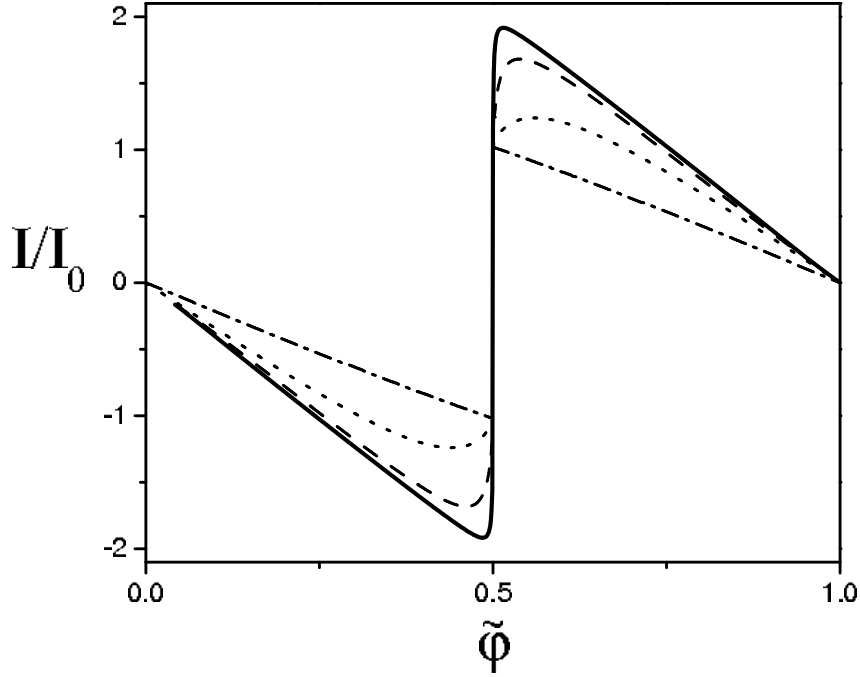}}
\caption{Persistent Current, in units of the persistent current of an isolated ring
$I_0$, as a function of $\tilde{\varphi}$ for different values of ring-wire
coupling,$\gamma=0.001$ (solid line),$\gamma=0.1$(dashed line),$\gamma=1$ (dotted line)
and $\gamma=100$ (dash-dot line).} \label{fig5}
\end{figure}

\begin{figure}[]
\vspace{2cm}
\centerline{\includegraphics[angle=-90,scale=0.5]{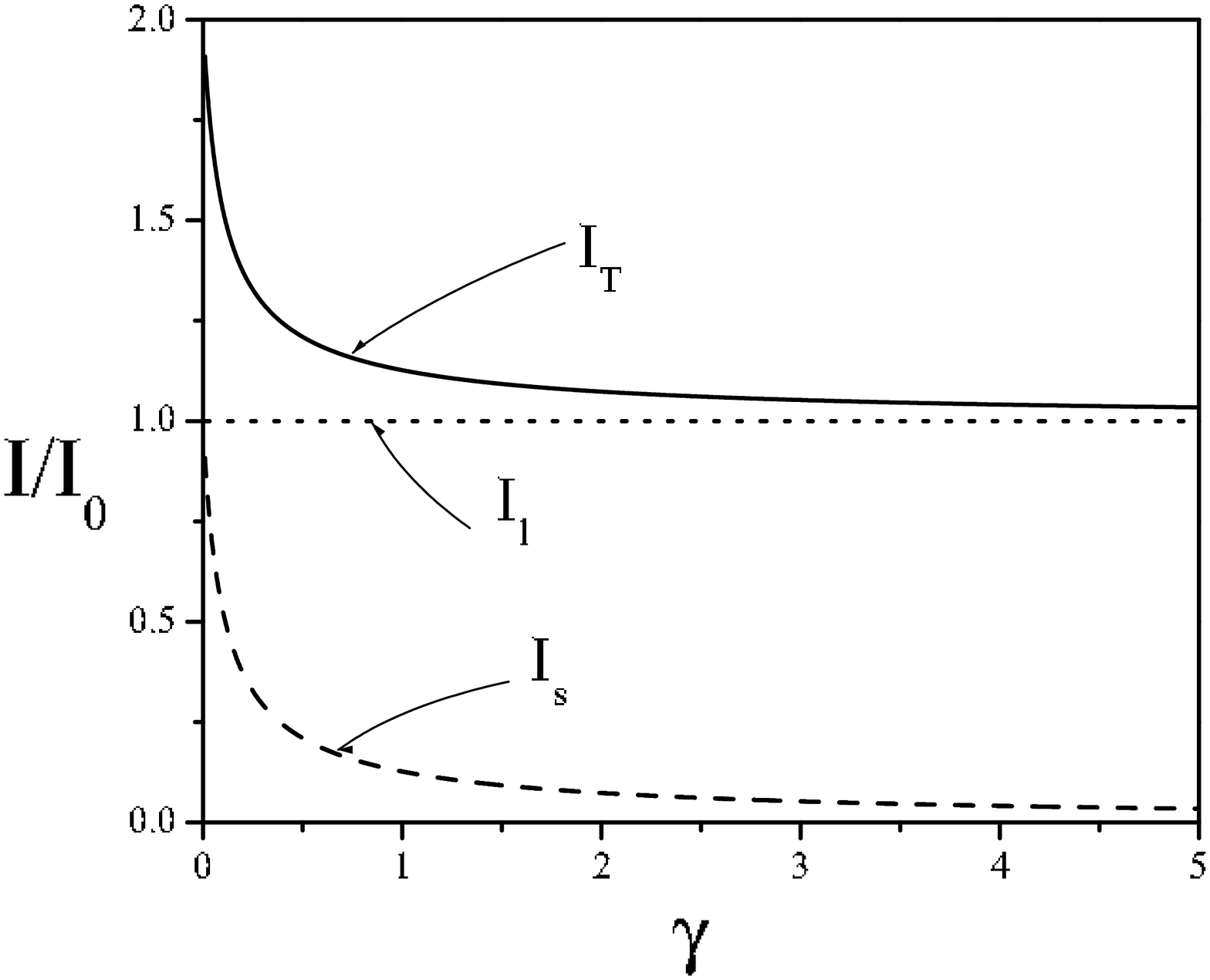}}
\caption{Amplitude of the persistent current(solid line) and $I_s$
(dashed line) and $I_b$(dotted line) as a function of $\gamma$
measured in units of $v$.} \label{fig6}
\end{figure}

\begin{figure}[h!]
\vspace{2cm}
\centerline{\includegraphics[angle=-90,scale=0.5]{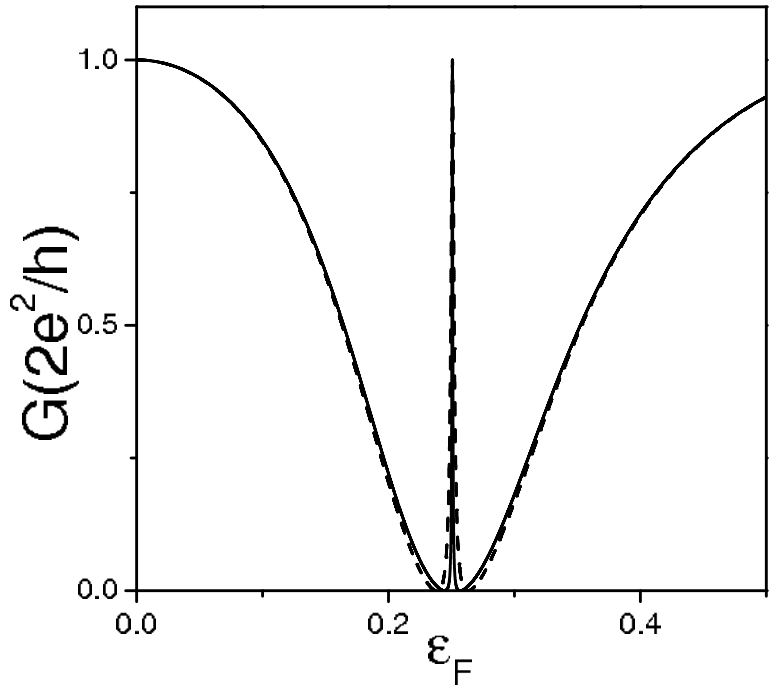}}
\caption{Linear conductance as a function of $\varepsilon_F$ (in
units of $v$) for $\delta\varphi=0.01$ (solid line) and $\delta
\varphi=1.99$ (dashed line).} \label{fig7}
\end{figure}

Now we examine the cases for  $\delta \varphi$ taking values close to an integer number.
For these cases $y \approx 0$ and the conductance can be written approximately as a
superposition of a broad Fano line shape and a narrow Breit-Wigner line shape around
$x=0$ with widths $\Gamma_{+}=2\gamma$ and $\Gamma_{-}=y^2/(2\gamma\beta)$ respectively.
This is,
\begin{equation}
\mathcal{G}\approx\frac{2e^2}{h}\big
(\frac{x^2}{x^2+\Gamma_{+}^2}+\frac{\Gamma_{-}^2}{x^2+\Gamma_{-}^2}\big).
\end{equation}
\noindent This expression clearly shows the superposition of  short and long living
states developed in the rings. Figure \ref{fig7} displays one of the narrow resonances of
conductance for $\tilde{\varphi}=0.6$, $\delta\varphi=0.01$ (solid line) and
$\delta\varphi=1.99$ (dashed line).

Similarly, the density of states and the persistent current
density can be written near $y=0$  in terms of contributions
coming from the two kind of states,

\begin{eqnarray}
\rho_{\alpha}(\omega)&\approx&
\frac{N\gamma}{sin(ka)^2}\frac{(1-\cos(Nka)\cos(\pi(\tilde{\varphi}\pm
\delta \varphi))-(2v\beta/N)(x\mp y))(x\pm
2y)x}{(x^2-y^2)^2+\Gamma_{+}^2x^2}\nonumber \\
&+&2vN\beta\frac{\Gamma_{-}}{x^2+\Gamma_{-}^2}
\end{eqnarray}

\begin{eqnarray}
 j_{\alpha}\approx
-\frac{e\sin(\pi(\tilde{\varphi}\pm \delta \varphi))}{h }\big
(\frac{x(x\pm2y)\Gamma_{+}}{(x^2-y^2)^2+\Gamma_{+}^2x^2}
+\frac{\Gamma_{-}}{x^2+\Gamma_{-}^2}\big )\
\end{eqnarray}

\noindent In Fig. 8 we show an example  of  the DOS (Fig. 8(a))
and the current density (Fig. 8 (b))as a function of the energy
for nearly equals magnetic fluxes, $\delta\varphi=0.01$, threading
the upper ring (solid line) or the lower ring (dashed line). The
coupling energy parameter is  $\gamma=0.001$. In Fig. 9 it is
shown the total persistent current as a function of the total of
magnetic flux for different values of $\gamma$ in the regime of
nearly equals magnetic fluxes $\delta\varphi =0.001$.

\begin{figure}[]
\centerline{\includegraphics[angle=-90,
scale=0.5]{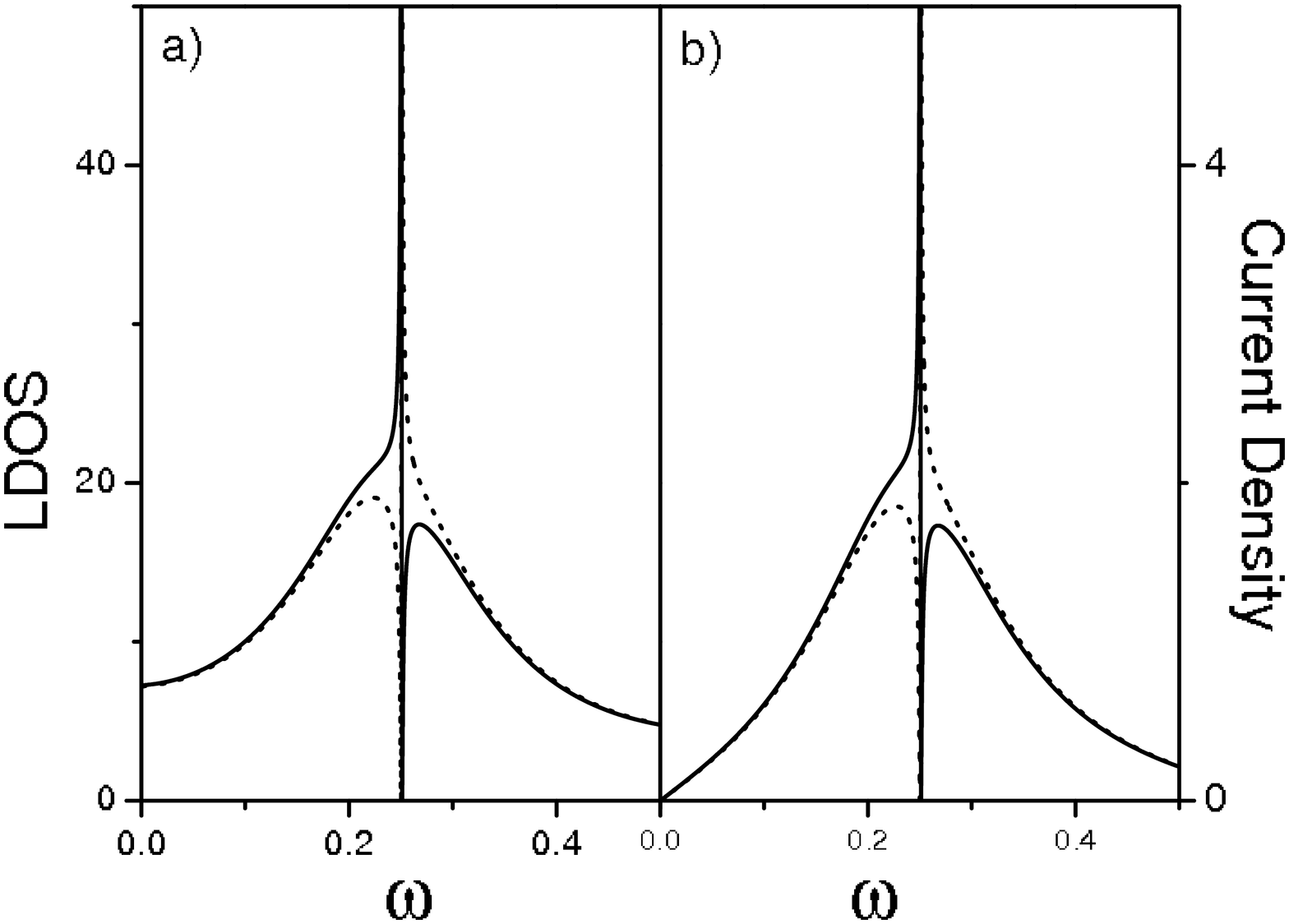}} \caption{a) DOS, b) current density
in the upper ring (solid line) and lower ring (dashed line) for
$\delta\varphi=0.01$ as a function of the energy (in units of
$v$).} \label{fig8}
\end{figure}

\begin{figure}[]
\centerline{\includegraphics[angle=-90, scale=0.5]{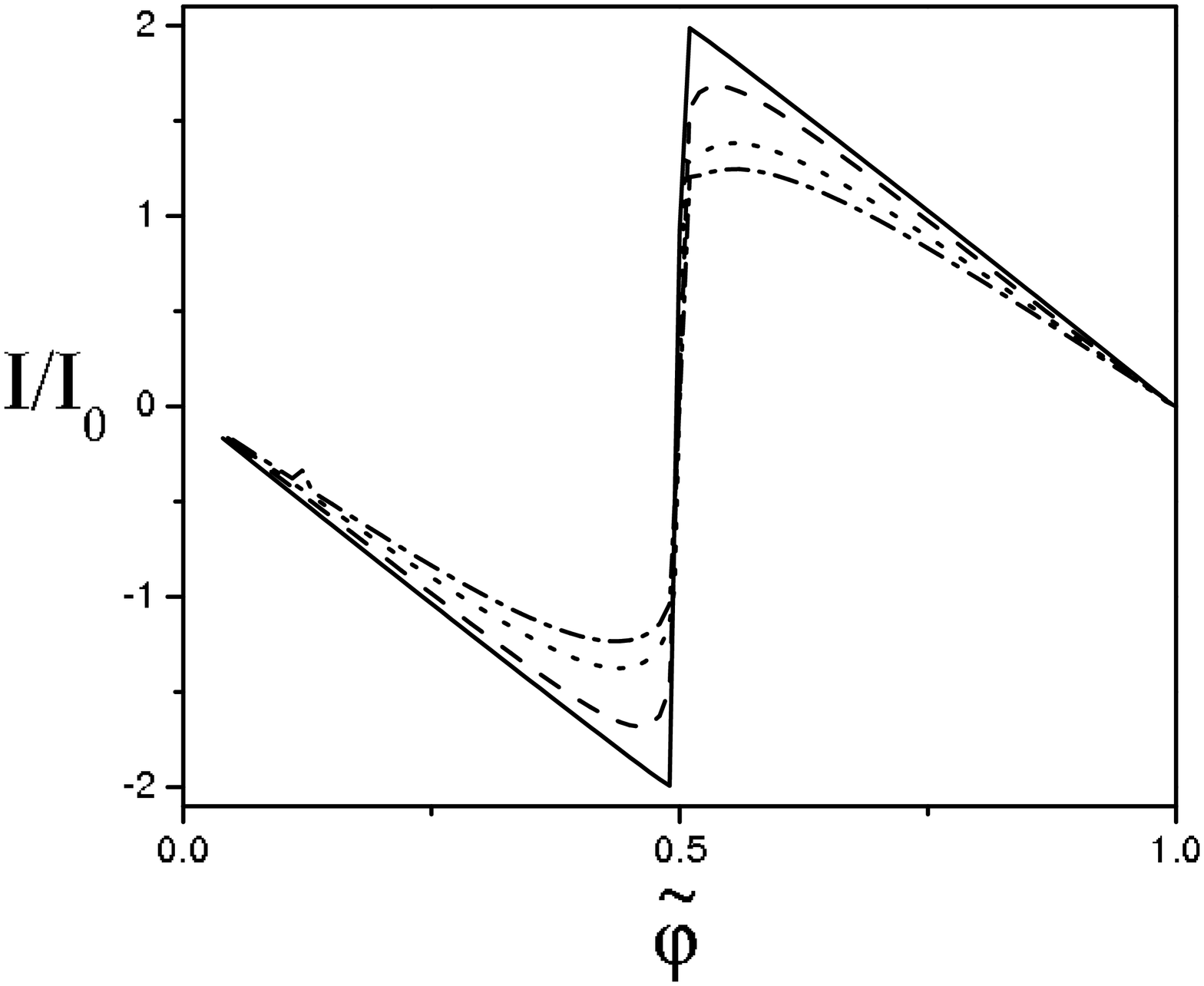}} \caption{Persistent
Current as a function of the magnetic flux for $\delta\varphi =0.001$ for different
values of coupling parameter, $\gamma=0.0001$ (solid line), $\gamma=0.001$ (dashed
line),$\gamma=0.5$ (dotted line), $\gamma=1$ (dash-dot line)} \label{fig9}
\end{figure}

The apparition of quasi-bound states in the spectrum of the system is a consequence of
the mixing of the levels of both rings which are coupled indirectly through the continuum
of states in the wire.  A similar effect was discussed recently in a system with a ring
coupled to a reservoir by Wunsch et al. in ref. [\onlinecite{wunsch}]. They relate this
kind of collective states with the Dicke effect in optics. The Dicke effect in optics
takes place in the spontaneous emission of a pair of atoms radiating a photon with a wave
length much larger than the separation between them. \cite{dicke} The luminescence
spectrum is characterized by a narrow and a broad peak, associated with long and
short-lived states, respectively.

\section{Summary}
We have investigated the persistent current, conductance and
density of states in a system of two side quantum rings attached
to a quantum wire in the presence of magnetic fluxes threading the
rings. In the regime of nearly degenerated isolated rings
spectrum, two kind of collective states are developed in the
system. States strongly coupled to the wire with an effective
coupling twice the coupling of a single ring, and strongly
localized states. We found that when either the flux difference or
the sum of the fluxes are close to zero or close to integer
multiples of a quantum of flux both, DOS and PCD show sharp and
broad peaks around the energy corresponding to the eigenenergies
of the isolated rings. In the limit when the isolated rings
spectrum is completely degenerated bound states are formed in the
rings leading to a magnification of the persistent currents in the
rings. We also found that the persistent current keeps a large
amplitude even for strong ring-wire coupling. This is a purely
quantum effect, and it is related to the formation of the
collective states in the double quantum-ring system.

\section*{ACKNOWLEDGMENTS}

P.\ A.\ O.\ and M.\ P.\ would like to thank financial support from
Milenio ICM P02-054-F and FONDECYT under grant 1020269 and
1010429.

\end{document}